\documentclass[10pt,preprint]{revtex4-1}
\usepackage{graphicx}
\usepackage{amsmath}
\usepackage{amssymb}
\usepackage{color}
\begin{document}


\title{Scaling law of high harmonic generation in the framework of photon channel}

\author{Liang Li,$^{1}$ Pengfei Lan$^{1}$}
\email{pengfeilan@hust.edu.cn}
\author{Lixin He$^{1}$}
\author{Xiaosong Zhu$^{1}$}
\email{zhuxiaosong@hust.edu.cn}
\author{Jing Chen$^{2,3}$}
\email{chen_jing@iapcm.ac.cn}
\author{Peixiang Lu$^{1,4}$}
\email{lupeixiang@hust.edu.cn}



\affiliation{%
 $^1$School of Physics and Wuhan National Laboratory for Optoelectronics, Huazhong University of Science and Technology, Wuhan 430074, China\\
 $^2$ HEDPS, Center for Applied Physics and Technology, Peking University, Beijing 100084, China \\
 $^3$ Institute of Applied Physics and Computational Mathematics, P. O. Box 8009, Beijing 100088, China\\
$^4$Laboratory of Optical Information Technology, Wuhan Institute of Technology, Wuhan 430205, China
}%

\date{\today}

\begin{abstract}
Photon channel perspective on
high harmonic generation (HHG) is proposed by quantizing both
the driving laser and high harmonics. It is shown that the HHG yield can be expressed as a sum of the contribution of all the photon channels.
From this perspective, the contribution of a specific photon channel follows a simple scaling law and the competition between the channels is well interpreted. Our prediction is shown to be in good agreement with the simulations by solving the time-dependent Schr\"{o}dinger equation. It also can well explains the experimental results of the HHG in the noncollinear two-color field and bicicular laser field.
\end{abstract}                         
\maketitle

High harmonic generation (HHG) is a highly nonlinear process in the interaction between the atom (or molecule) and the intense laser field. Single or trains of attosecond pulses can be generated by coherently synthesizing a series of high harmonics \cite{Farkas1992,Christov1997,Paul2001,Hentschel2001}, which enables us to steer and probe the nuclear and electronic dynamics in an unprecedentedly fast time scale \cite{Corkum2007,Salieres2012,Baker2006}.

Many theoretical methods have been developed to investigate the HHG. By solving the time-dependent Schr\"{o}dinger equation (TDSE), one can reproduce the HHG. However, the underlying physics can not be straightforward revealed since the rich information is encoded in the electronic wave function. Then many other approaches such as Lewenstein model \cite{Lewenstein1994}, quantum orbits (QO) theory \cite{Becker2002,Salieres2001,Sansone2004}, quantitatively rescattering (QRS) theory \cite{CDL2010,Morishita2008,CDL2009} and factorization methods \cite{Frolov2007,Ivanov2009} have been developed. With these theories, the HHG process can be quantitatively described in terms of the electron trajectories (or quantum orbits \cite{Becker2002}), and lots of experimental results can be well interpreted, e.g., the cut-off law and the long or short trajectory.

All the theories mentioned above treat the driving laser and the generated high harmonics as classical oscillatory electromagnetic waves. Although it has proven quite successful in describing several facets of HHG, it is not convenient to explain the quantized photon properties of HHG. However, the theory treating the driving laser and the generated high harmonics as photons is scarce \cite{Becker1997,Gaolh2000,Wangbb2007}. Recently, HHG in the noncollinear two-color laser field \cite{Bertrand2011,Daniel2015} and non-pure vortex beam \cite{Laura2016} are drawing increasing attentions in both fundamental studies and applications. To explain the complicated features of the generated high harmonics, the concept of photon channel was employed. In this case, one can intuitively understand the HHG with photon channels that an atom absorbs and emits specific numbers of photons based on the selection rules or conservation laws \cite{Bertrand2011,Daniel2015,Laura2016,Fleischer2014}. However, the photon channel is only phenomenologically applied. The HHG theory from photon channel perspective is far from being quantitatively formulated and the underlying physics is still not well understood.

In this Letter, we propose a photon channel perspective on HHG by quantizing the electromagnetic field. From this perspective, the HHG emission yield can be expressed in terms of the photon channels. The contribution of a specific photon channel to HHG follows a simple analytical formula. Then the complicated photon-like features of HHG in multi-mode field can be well described, such as the photon channel competition and the scaling law of HHG yield.

We consider the HHG in a two-color field. The laser intensities, frequencies and wave vectors are denoted as $I_1$, $I_2$, $\omega_{1}$, $\omega_{2}$ and $\mathbf{k}_{1}$, $\mathbf{k}_{2}$, respectively. The frequency and wave vector of the emitted high harmonics are denoted as $\Omega$ and $\mathbf{k}^{\prime}$, respectively. The two-color field and the high harmonics are both quantized.
Vacuum polarization and other relativistic effects for the electron are ignored. The Hamiltonian of this atom-radiation system is:
\begin{equation}
  H = H_{0}+H_{p}+V_{L}
\end{equation}
where $H_{0}=\hat{T}+\hat{V}_{c}$ and $H_{p}=\omega_{1} \hat{N_{1}}+\omega_{2} \hat{N_{2}}+\Omega \hat{N_{\Omega}}$ are Hamiltonian of the field-free atom and photon, respectively. $\hat{T}$ and $\hat{V}_{c}$ are the kinetic energy and potential of the electron. $\hat{N_{1}}=(a_{1}^{\dag}a_{1}+a_{1}a_{1}^{\dag})/2$, $\hat{N_{2}}=(a_{2}^{\dag}a_{2}+a_{2}a_{2}^{\dag})/2$, $\hat{N_{\Omega}}=((a^{\prime})^{\dag}a^{\prime}+a^{\prime}(a^{\prime})^{\dag})/2$ are the photon number operators of the two-color field and the harmonic photon mode, respectively. $a$ and $a^{\dag}$ are the annihilation and creation operators. $V_{L} = -\textbf{d} \cdot (\textbf{E}_1+\textbf{E}_2+\textbf{E}^{\prime}) $ is the electron-photon interaction. The electric fields $\textbf{E}_m$ ($m=1, 2$) and $\textbf{E}^{\prime}$ for the driving laser and the generated harmonics can be expressed as $\textbf{E}_m= i g_{m}(\hat{\textbf{$\epsilon$}}_{m}a_{m}e^{i\textbf{k}_{m}\cdot\textbf{r}}-c.c.)$ and $\textbf{E}^{\prime}=i g^{\prime}(\hat{\textbf{$\epsilon$}}^{\prime}a^{\prime}e^{i\textbf{k}^{\prime}\cdot\textbf{r}}-c.c.)$.
$g_{m}=(2\omega_{m}/V)^{1/2}$, $g^{\prime}=(2\Omega/V^{\prime})^{1/2}$, where $V$ and $V^{\prime}$ are the normalization volumes of the driving laser and high harmonics. In large photon-number limit $g\sqrt{N}\rightarrow \sqrt{\frac{I}{2}}$ \cite{Gaolh2000}. $\hat{\textbf{$\epsilon$}}_{m}=\hat{\textbf{$\epsilon$}}_{x}\cos(\theta_{m})+i\hat{\textbf{$\epsilon$}}_{y}\sin(\theta_{m})$ and $\hat{\textbf{$\epsilon$}}^{\prime}=\hat{\textbf{$\epsilon$}}_{x}\cos(\theta^{\prime})+i\hat{\textbf{$\epsilon$}}_{y}\sin(\theta^{\prime})$ are the transverse polarization. The long-wavelength approximation ($\lambda\gg r_{electron}$) is considered in this work, i.e. $e^{i\textbf{k}_{m}\cdot r}\approx 1$, $e^{i\textbf{k}^{\prime} \cdot r}\approx 1$ and the electric field is independent of $r$.

To clarify interaction between the atom and the photons, the Hamiltonian is rewritten in Interaction picture
\begin{eqnarray}\label{Eq:HI}
  H_{I}(t) &=& e^{i(H_{0}+H_{p})t}(-d\cdot(\textbf{E}_1+\textbf{E}_2+\textbf{E}^{\prime}))e^{-i(H_{0}+H_{p})t} \nonumber\\\emph{\emph{}}
  &=& -D(t)\cdot (\varepsilon_{1}(t)+\varepsilon_{2}(t)+\varepsilon^{\prime}(t))
\end{eqnarray}
where $D(t)=e^{iH_{0}t}de^{-iH_{0}t}$, $\varepsilon_{1,2}(t)= i g_{1,2}(\hat{\epsilon}_{1,2}a_{1,2}e^{-i \omega_{1,2}t}-c.c.)$ and $\varepsilon^{\prime}= i g^{\prime}(\hat{\epsilon}^{\prime}a^{\prime}e^{-i\Omega t}-c.c.)$. Then, the transition matrix element between two states $|i\rangle$ and $|f\rangle$ is
\begin{eqnarray}\label{Eq:transition}
  A(i\rightarrow f) &=& \langle f|e^{-i(H_{0}+H_{p})t}U_{I}(t,t_{0})|i\rangle,
\end{eqnarray}
where $U_{I}(t,t_{0})$ is the time-evolution operator in Interaction picture. Using the Dyson equation, $U_{I}$ can be expressed as \cite{Dyson1949}
\begin{eqnarray}\label{Eq:dyson}
  U_{I}(t,t_{0}) &=& \sum^{\infty}_{n=0}U_{In}(t,t_{0}), \\
  U_{In}(t,t_{0}) &=& (\frac{1}{i})^{n}\int^{t}_{t_{0}}dt_{1}...\int^{t_{n-1}}_{t_{0}}dt_{n}H_{I}(t_{1})...H_{I}(t_{n}).
\end{eqnarray}
Substituting Eqs. \ref{Eq:HI} and \ref{Eq:dyson} into Eq. \ref{Eq:transition}, we have
\begin{eqnarray}\label{Eq:a}
  A(i\rightarrow f) &=& \sum_{n=0}^{\infty}\langle f|e^{-i(H_{0}+H_{p})t}(\frac{1}{i})^{n}\int^{t}_{t_{0}}dt_{1}...\int^{t_{n-1}}_{t_{0}}dt_{n}H_{I}(t_{1})...H_{I}(t_{n})|i\rangle, \nonumber\\\emph{\emph{}}
  &=& \sum_{n=0}^{\infty}\sum_{\omega^{t_{1}},\omega^{t_{2}},...,\omega^{t_{n}}}A_{n}(\omega^{t_{1}},\omega^{t_{2}},...,\omega^{t_{n}})
\end{eqnarray}
\begin{eqnarray}\label{Eq:an}
  A_{n}(\omega^{t_{1}},\omega^{t_{2}},\omega^{t_{3}},...,\omega^{t_{n}}) =\int_{t_{0}}^{t}dt_{1}\int_{t_{0}}^{t_{1}}dt_{2}...\int_{t_{0}}^{t_{n-1}}dt_{n}\ \ \ \ \ \ \ \ \ \ \ \ \ \ \ \ \ \ \ \ \ \ \ \ \ \ \ \ \   \nonumber\\\emph{\emph{}}
  \times\langle f|e^{-i(H_{0}+H_{p})t}(-D(t_{1})\varepsilon(\omega^{t_{1}}))(-D(t_{2})\varepsilon(\omega^{t_{2}}))...(-D(t_{n})\varepsilon(\omega^{t_{n}}))|i\rangle \ \ \ \ \ \ \ \ \ \ \ \ \ \ \ \
\end{eqnarray}
where $\varepsilon(\omega_{1,2}) = \sqrt{\frac{2\omega_{1,2}}{V}}\hat{\epsilon}_{1,2}a_{1,2}e^{-i\omega_{1,2}t}$, $\varepsilon(-\omega_{1,2}) = -\sqrt{\frac{2\omega_{1,2}}{V}}\hat{\epsilon}_{1,2}^{\ast}a^{\dag}_{1,2}e^{i\omega_{1,2}t}$,
$\varepsilon(\Omega) = \sqrt{\frac{2\Omega}{V^{\prime}}}\hat{\epsilon}^{\prime}a^{\prime}e^{-i\Omega t}$ and $\varepsilon(-\Omega) = -\sqrt{\frac{2\Omega}{V^{\prime}}}(\hat{\epsilon}^{\prime})^{\ast}(a^{\prime})^{\dag}e^{i\Omega t}$. $\omega^{t_{i}}$ is the frequency of the photon absorbed/emitted at time $t_{i}$, i.e. $\omega^{t_{i}}=\pm\omega_{1}$ , $\pm\omega_{2}$ or $\pm\Omega$ (``$+$'' denotes absorption and ``$-$'' denotes emission).

In our model, the HHG process is described by the transition from the initial state to the final state of the atom via the interaction with the driving photon field. As in the previous well-known models \cite{Lewenstein1994,Frolov2007,Ishikawa2009,Frolov2010,vveden2016}, the initial and final states are the ground state of the atom. All the transitions between the excited bound states and continuous states are neglected and the depletion of the atom is not taken into account. In details, the initial and final states in our model are written as $|i\rangle=|\phi_{0},N_{1i},N_{2i},0\rangle=\phi_{0}\bigotimes|N_{1i}\rangle$
$\bigotimes|N_{2i}\rangle\bigotimes|0\rangle^{\prime}$ and $|f\rangle=|\phi_{0},N_{1f},N_{2f},1\rangle=\phi_{0}\bigotimes|N_{1f}\rangle\bigotimes|N_{2f}\rangle\bigotimes|1\rangle^{\prime}$, which are the eigenstates of the Hamiltonian $H_{0}+H_{P}$ with eigenenergies $E_{i}=-E_{0}+(N_{1i}+\frac{1}{2})\omega_{1}+(N_{2i}+\frac{1}{2})\omega_{2}+\frac{1}{2}\Omega$ and $E_{f}=-E_{0}+(N_{1f}+\frac{1}{2})\omega_{1}+(N_{2f}+\frac{1}{2})\omega_{2}+\frac{3}{2}\Omega$, respectively. $\phi_{0}$ is the ground-state wave-function of the atomic electron with binding energy $E_{0}$. $|N_{1i}\rangle$, $|N_{2i}\rangle$, $|N_{1f}\rangle$ and $|N_{2f}\rangle$ are the Fock states of the laser modes with photon number $N_{1i}$, $N_{2i}$, $N_{1f}$ and $N_{2f}$. $|0\rangle^{\prime}$ and $|1\rangle^{\prime}$ are the Fock states of the high harmonic. According to the law of conservation of energy, only the terms $A_{n}(-\Omega;\omega^{t_{2}},\omega^{t_{3}},...,\omega^{t_{n}})$ satisfying $\Omega=\omega^{t_{2}}+\omega^{t_{3}}+...+\omega^{t_{n}}$ contribute to the harmonic $\Omega$. Finally, by using $a|N\rangle=\sqrt{N}|N-1\rangle$ and $a^{\dag}|N\rangle=\sqrt{N+1}|N+1\rangle$, the emission rate of the harmonic $\Omega$ can be expressed as (see Sec. A in the supplementary material \cite{SM})
\begin{eqnarray}\label{Eq:P}
  P(\Omega) &=& |A(i\rightarrow f)|^{2} \nonumber\\\emph{\emph{}}
  &=& |\sum_{n=0}^{\infty}\sum_{\omega^{t_{2}},...,\omega^{t_{n}}}A_{n}(-\Omega;\omega^{t_{2}},...,\omega^{t_{n}})\delta(\omega^{t_{2}}+...+\omega^{t_{n}}-\Omega)|^{2} \nonumber\\\emph{\emph{}}
  &=& |\sum_{n=0}^{\infty}\sum_{\omega^{t_{2}},...,\omega^{t_{n}}} \sigma_{0}^{\frac{1}{2}}(-\Omega;\omega^{t_{2}},...,\omega^{t_{n}})p^{\frac{1}{2}}(|\omega^{t_{2}}|)...p^{\frac{1}{2}}(|\omega^{t_{n}}|)\delta(\omega^{t_{2}}+...+\omega^{t_{n}}-\Omega)|^{2}
\end{eqnarray}
\begin{align}
\nonumber \sigma_{0}^{\frac{1}{2}}(-\Omega;\omega^{t_{2}},...,\omega^{t_{n}})=\int_{t_{0}}^{t}dt_{1}...\int_{t_{0}}^{t_{n-1}}dt_{n}\langle\phi_{0},N_{1f},N_{2f},1|D(t_{1})(\hat{\epsilon}^{\prime})^\ast\sqrt{\frac{2\Omega}{V^\prime}}e^{i\Omega t_{1}}(a^{\prime})^{\dagger} \\
\times (-1)^{n-1}(D(t_{2})\hat{\epsilon}^{t_{2}}\sqrt{\frac{I}{2}}e^{-i\omega^{t_{2}}t_{2}})...(D(t_{n})\hat{\epsilon}^{t_{n}}\sqrt{\frac{I}{2}}e^{-i\omega^{t_{n}}t_{n}})|\phi_{0},N_{1f},N_{2f},0\rangle
\end{align}
where $\hat{\epsilon}^{t_{i}}=\hat{\epsilon}_{m}$ for $\omega^{t_{i}}=\omega_{m}$ and $\hat{\epsilon}^{t_{i}}=-\hat{\epsilon}_{m}^{\ast}$ for $\omega^{t_{i}}=-\omega_{m}$ ($m=1,2$). $p(\pm\omega_{m})=p_{m}=\frac{I_{m}}{I}$ $(m=1,2)$ are the ratios of the intensity. $\sigma_{0}^{\frac{1}{2}}(-\Omega;\omega^{t_{2}},\omega^{t_{3}},...,\omega^{t_{n}})$ describes the ability of emitting a harmonic photon $\Omega$ via a quantum path of absorbing a series photons $\omega^{t_{n}}$, $\omega^{t_{n-1}}$... $\omega^{t_{2}}$.

To establish the link between our model and the observable quantity, we introduce the photon channel that is the sum of all the quantum paths involving the same net number $(n_{1},n_{2})$ of the two color photons. This photon channel corresponds to the high harmonic $\Omega=n_1\omega_1+n_2\omega_2$ that are observable in experiment. From the photon channel perspective, the emission rate of the harmonic $\Omega$ can be expressed as:
\begin{eqnarray}\label{Eq:P2}
  P(\Omega) &=& \sum_{n_{1},n_{2}}P(\Omega(n_{1},n_{2}))\delta(n_{1}\omega_{1}+n_{2}\omega_{2}-\Omega) \nonumber\\\emph{\emph{}}
  &=& \sum_{n_{1},n_{2}}\sigma(n_{1},n_{2})p_{1}^{|n_{1}|}p_{2}^{|n_{2}|}\delta(n_{1}\omega_{1}+n_{2}\omega_{2}-\Omega)
\end{eqnarray}
where $\sigma(n_{1},n_{2})=|\sum_{[n_{1},n_{2}]}\sigma_{0}^{\frac{1}{2}}(\Omega;\omega^{t_{2}},\omega^{t_{3}},...,\omega^{t_{n}})e^{i\varphi_{[n_{1},n_{2}]}}|^{2}$. $[n_{1},n_{2}]$ indicates an arrangement of net $n_{1}$ $\omega_{1}$-photons and net $n_{2}$ $\omega_{2}$-photons. Note that the net $n_m$ ($m=1,2$) photon absorption process may involve absorption of $(n_m+1)$ photons and emission of $1$ photon, absorption of $(n_m+2)$ photons and emission of $2$ photons, and so on. We can deal with these terms by using the ansatz: the sum of all the terms involving extra absorption and emission of photons in the summation only give rise to a phase factor (see Sec. B in \cite{SM}). Furthermore, we introduce the permutation symmetry to the parameter $\sigma_{0}$ with the standard method \cite{Nonlinear} and then $\sigma_{0}$ for different quantum paths in a specific photon channel become equal. Therefore we can obtain $\sigma(n_{1},n_{2})=C^{|n_{1}|}_{|n_{1}|+|n_{2}|}\sigma_{0}(n_{1},n_{2})$, and then $P(\Omega(n_{1},n_{2}))=\sigma_{0}(n_{1},n_{2})C_{|n_{1}|+|n_{2}|}^{|n_{1}|}p_{1}^{|n_{1}|}p_{2}^{|n_{2}|}$, where $\sigma_0(n_1,n_2)$ is the abbreviation of $\sigma_{0}(\Omega;\omega^{t_{2}},\omega^{t_{3}},...,\omega^{t_{n}})$. Note that this formula can be separated to two terms: the term $\sigma_{0}$ describes the characteristic structure of the high harmonic spectra and the term $C_{|n_{1}|+|n_{2}|}^{|n_{1}|}p_{1}^{|n_{1}|}p_{2}^{|n_{2}|}$ corresponds to the weight of the photon channel, which describes the channel competition and power scaling of a specific harmonic (see the discussion below). Interestingly, the second term has the similar behavior to the nonlinear optical wave mixing in the perturbation regime \cite{Nonlinear}. It indicates that, although HHG is a highly nonperturbative process, the power scaling of its photon channel still follows a perturbative way. A recent experiment has demonstrated this property in the case of $I_{2}\ll I_{1}$ \cite{Bertrand2011}. However, this phenomenon is only explained using a phenomenological scaling of $I_{2}^{n_{2}}$ in Ref. \cite{Bertrand2011}. Here we provide a quantitative model for understanding the perterbative property of the photon channel. As shown below, our model will retrieve the same scaling law $I_{2}^{n_{2}}$ as in \cite{Bertrand2011} if $I_{2}\ll I_{1}$. More importantly, our formula still works for stronger $I_2$ where $I_{2}^{n_{2}}$ scaling law fails. It therefore provide a more complete and comprehensive photon-channel perspective of HHG.



\begin{figure}[!t]
  \includegraphics[width=10cm]{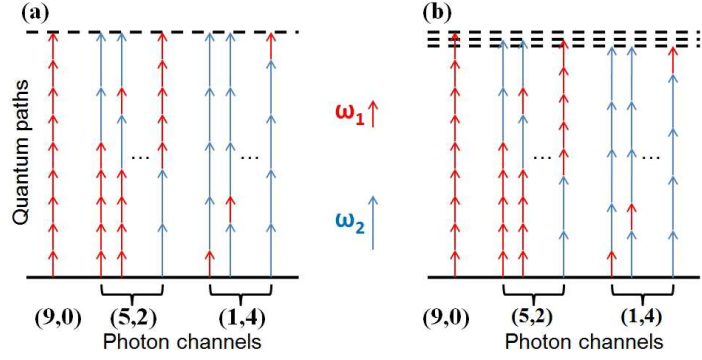}
  \caption{The sketches of the quantum paths and photon channels of HHG in the two-color field. A photon channel contains many quantum paths as denoted in the brace. (a) The sketch for harmonic $\Omega=9\omega_{1}$ with $\omega_{1}:\omega_{2}=1:2$. In this case, the photon channels $\Omega(9,0)$, $\Omega(5,2)$, $\Omega(1,4)$ are degenerate. (b) The sketch for harmonics $\Omega=9\omega_{1}, 8.8\omega_{1}$ and $8.6\omega_{1}$ with $\omega_{1}:\omega_{2}=1:1.9$.}\label{fig1}
\end{figure}

To validate our model, we numerically solve the three dimensional TDSE \cite{Mpro1997} in a two-color filed. Incommensurate frequencies $\omega_{1}:\omega_{2}=1:1.9$ instead of $1:2$ are applied to identify the photon channels \cite{Fleischer2014}. As shown in Fig. \ref{fig1}, there are many photon channels, such as $\Omega(9,0), \Omega(5,2), \Omega(1,4)$ and so on, contributing to one harmonic $\Omega=9\omega_{1}$ when using a two-color field with frequencies $1:2$. By using a two-color field with incommensurate frequencies $1:1.9$, the degenerate channels become distinguishable. The linearly polarized 800-nm ($\omega_{1}$) and 421-nm ($\omega_{2}=1.9\omega_{1}$) fields are adopted in the simulation and the target atom is hydrogen. The laser field is turned on linearly over the first 10 optical cycles and is kept constant for another 110 optical cycles. We keep the total intensity $I=I_{1}+I_{2}$ of the two-color field constant ($0.2\times10^{14}$W/cm$^{2}$) and vary the ratio $p_{2}=I_{2}/I$.

\begin{figure}[!t]
  \centering
  \includegraphics[width=10cm]{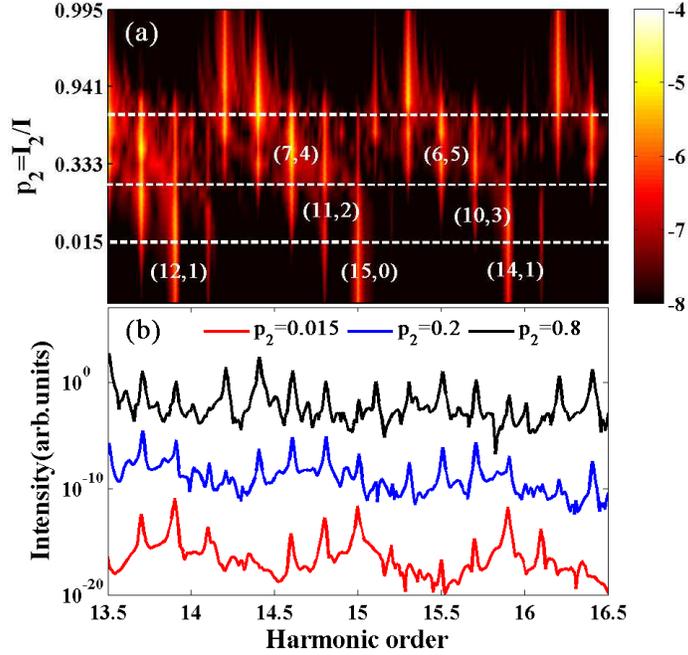}
  \caption{(a) The HHG spectra for different $p_{2}$. Each photon channel for $\Omega=(n_{1}+1.9n_{2})\omega_{1}$ is labeled as $(n_{1},n_{2})$. The colorbar denotes the harmonic yield in logarithmic scale. (b) The HHG spectra for $p_{2}=0.015,0.2$ and $0.8$. The blue and black curves are shifted vertically by multiplying a factor of $10^{6}$ and $10^{12}$, respectively.}\label{fig2}
\end{figure}

Figure \ref{fig2} shows the high harmonic spectra obtained with TDSE for different $p_{2}$. The photon channels, e.g., $\Omega(15,0)$, $\Omega(11,2)$ and $\Omega(7,4)$, can be clearly identified. For a small $p_{2}(=0.015)$, the channels with small $n_{2}$ are dominant, e.g. $\Omega(12,1)$, $\Omega(15,0)$ and $\Omega(14,1)$. This agrees well with the prediction of our model (see Eq. \ref{Eq:P2}) that $p_{2}^{|n_{2}|}$ rapidly decreases with increasing $n_{2}$. One can clearly see from Fig. 2 that the dominant photon channel converts from smaller $n_2$ to larger $n_2$ with increasing $p_2$. For example, the photon channel $\Omega(15,0)$ converts to $\Omega(11,2)$, $\Omega(7,4)$ and so on. One can also see the channel competition around $p_2=0.2$, e.g. $\Omega(11,2)$ and $\Omega(7,4)$ are comparable for $p_2=0.2$. Such a complicated HHG spectrum due to the channel competition can be well predicted and explained with Eq. \ref{Eq:P2}. To evaluate the channel competition between $\Omega_1(n_1, n_2)$ and $\Omega_2(n'_1, n'_2)$, we introduce the ratio $\gamma(\Omega_{1},\Omega_{2})=P(\Omega_{1}(n_{1},n_{2}))/P(\Omega_{2}(n'_{1},n'_{2}))=\gamma_{0} C_{|n_{1}|+|n_{2}|}^{|n_{1}|}/C_{|n'_{1}|+|n'_{2}|}^{|n'_{1}|}(1-p_{2})^{|n_{1}|-|n'_{1}|}p_{2}^{|n_{2}|-|n'_{2}|}$, where $\gamma_{0}=\sigma_{0}(n_{1},n_{2})/\sigma_{0}(n_{1}^{\prime},n_{2}^{\prime})$. Here $\gamma_0$ can be obtained by solving the equation $\gamma_{0} C_{|n_{1}|+|n_{2}|}^{|n_{1}|}/C_{|n'_{1}|+|n'_{2}|}^{|n'_{1}|}(1-p_{2})^{|n_{1}|-|n'_{1}|}p_{2}^{|n_{2}|-|n'_{2}|}=1$, where the value of $p_2$ is determined according to $P(\Omega_{1}(n_{1},n_{2}))=P(\Omega_{2}(n'_{1},n'_{2}))$ in the TDSE simulation. Then, $\gamma(\Omega(15,0),\Omega(11,2))$ is calculated to be 10.1 for $p_2=0.2$, which predicts that the channel with smaller $n_2$ $\Omega(15,0)$ is dominant. For $p_2=0.8$, $\gamma(\Omega(7,4),\Omega(3,6))=0.04$ and the channel with larger $n_2$ $\Omega(3,6)$ is dominant.

\begin{figure}[!t]
  \includegraphics[width=10cm]{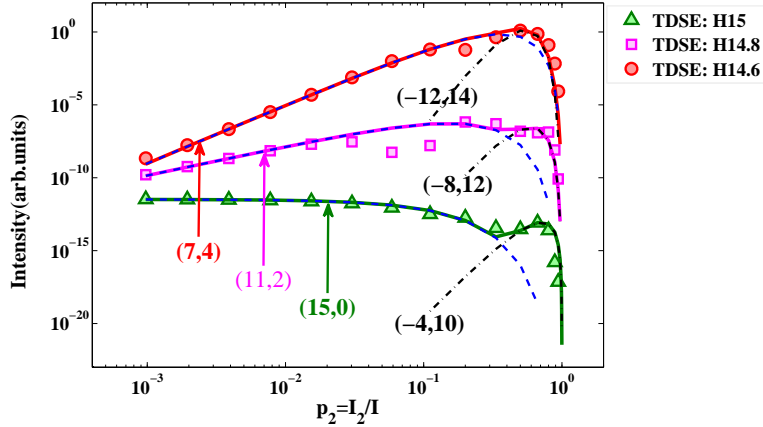}
  \caption{High harmonic yield as a function of $p_2$. The triangles, squares and circles represent the harmonic yields obtained with TDSE simulations. The dash and the dash-dot curves show the contributions of different photon channels (n$_1$, n$_2$) predicted with our model and the solid curves show the sum of them. The laser parameters are same as Fig. \ref{fig2}.}\label{fig3}
\end{figure}

Next we discuss the scaling law of the harmonic yield for a specific photon channel $\Omega(n_{1},n_{2})$. It should be noted that the degeneracy of the photon channels cannot be perfectly eliminated even by using the two-color field with incommensurate frequencies. For example, the same high harmonic $\Omega$ can be possibly contributed by both the photon channels $\Omega(n_{1},n_{2})$ and $\Omega(n_{1}-19,n_{2}+10)$. However, the harmonic $\Omega$ is usually dominated by only one photon channel since the ratio $\gamma=\gamma_0 C_{|n_{1}|+|n_{2}|}^{|n_{1}|}/C_{|n_{1}-19|+|n_{2}+10|}^{|n_{1}-19|} (1-p_{2})^{|n_{1}|-|n_{1}-19|}p_{2}^{|n_{2}|-|n_{2}+10|}$ is a function changing very fast with $p_{2}$ (either $\ll1$ or $\gg1$ for most values of $p_2$). Figure \ref{fig3} shows the high harmonic yield as a function of $p_{2}$ on the log-log scale. The triangles, squares and circles represent the yields of the harmonics $15\omega_{1}$, $14.8\omega_{1}$ and $14.6\omega_{1}$ obtained by numerically solving TDSE, respectively. The dash and dash-dot curves show the harmonic yields contributed by two degenerate photon channels ($\Omega(n_{1},n_{2})$ and $\Omega(n_{1}-19,n_{2}+10)$) predicted by our model and the solid curves show their sum. We take the yield of harmonic $15\omega_{1}$ as an example. Our model predicts that the channel $\Omega(15,0)$ is dominant for $p_{2}<0.3$ and the yield is proportional to $(1-p_{2})^{15}p_{2}^{0}$. When $p_{2}>0.3$, the dominant channel is converted from $\Omega(15,0)$ to $\Omega(-4,10)$ and the yield is proportional to $(1-p_{2})^{4}p_{2}^{10}$. These predicted scalings are in good agreement with the TDSE simulations. The same agreement is shown in Fig. \ref{fig3} for harmonics $14.8\omega_{1}$ and $14.6\omega_{1}$. In addition, one can find that the channel conversion is faster for larger $n_{1}$. This can be well explained with the ratio $\gamma$. For $\Omega(7,4)$ and $\Omega(-12,14)$ with smaller $n_1$, $\gamma=\gamma_0 C_{11}^{4}/C_{26}^{12}(1-p_{2})^{-5}p_{2}^{-10}$ has a minimum and changes slowly near $\gamma=1$, i.e. the conversion of dominant channel is slow. In contrast, for $\Omega(15,0)$ and $\Omega(-4,10)$ with larger $n_1$, $\gamma=\gamma_0C_{15}^{0}/C_{14}^{4}(1-p_{2})^{11}p_{2}^{-10}$ decreases monotonically with $p_{2}$ and has a big slope at $\gamma=1$, i.e., the conversion of dominant channel is faster.


\begin{figure}[!t]
  \includegraphics[width=10cm]{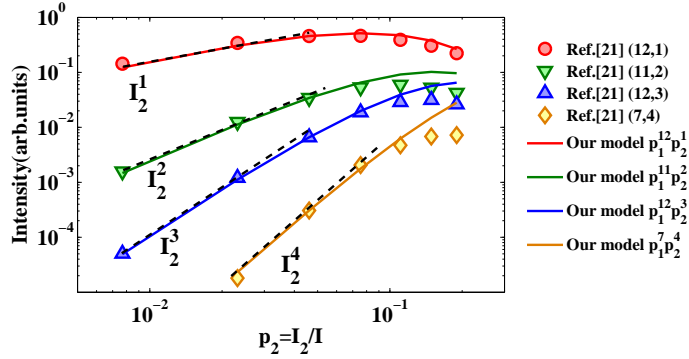}
  \caption{Comparisons between the HHG yields in the nonconlinear two-color field obtained from Ref. \cite{Bertrand2011} (dots) and our model (solid curves). The laser parameters are the same as those in Fig. 3(b) of \cite{Bertrand2011}.}\label{fig4}
\end{figure}
Our model can be generalized for other form of laser field rather than only for the collinear two-color field. For example, the emission rate in a noncollinear two-color field can also be obtained as the sum of photon channels by adopting the same procedure as above
\begin{eqnarray}\label{Eq:P3}
  P(\Omega,\mathbf{k}^{\prime}) &=& \sum_{n_{1},n_{2}}P(\Omega(n_{1},n_{2}))\delta(n_{1}\omega_{1}+n_{2}\omega_{2}-\Omega)\delta(n_{1}\mathbf{k}_{1}+n_{2}\mathbf{k}_{2}-\mathbf{k}^{\prime}) \nonumber\\\emph{\emph{}}
  &=& \sum_{n_{1},n_{2}}\sigma(n_{1},n_{2})p_{1}^{|n_{1}|}p_{2}^{|n_{2}|}\delta(n_{1}\omega_{1}+n_{2}\omega_{2}-\Omega)\delta(n_{1}\mathbf{k}_{1}+n_{2}\mathbf{k}_{2}-\mathbf{k}^{\prime})
\end{eqnarray}
To confirm this, we compare the predictions of our model with the previous work in \cite{Bertrand2011}.
We adopt the same laser intensity as in \cite{Bertrand2011}, where the intensity $I_{1}$ is fixed and $I_{2}$ is varied. Although the total laser intensity $I=I_1+I_2$ varies, we still have $p_{1}=\frac{I_{1}}{I_{1}+I_{2}}$ and $p_{2}=\frac{I_{2}}{I_{1}+I_{2}}$. Assuming that $I_2\ll I_1$ and the ground state depletion can be neglected in our calculation, $\sigma(n_{1},n_{2})$ is constant and then we can obtain the power scaling as $p_1^{|n_1|}p_2^{|n_2|}$. In Fig. \ref{fig4}, the harmonic yields obtained from \cite{Bertrand2011} and our model are shown as the dots and solid curves, respectively. When $p_{2}<0.04$, the results in \cite{Bertrand2011} follow the scalings $I_{2}^{1}$,
$I_{2}^{2}$ and $I_{2}^{3}$ for the channels $\Omega(12,1)$, $\Omega(11,2)$ and $\Omega(12,3)$, respectively. However, the harmonic yields in \cite{Bertrand2011} deviate significantly from the scaling $I_{2}^{n_{2}}$ for $p_2>0.04$. They become saturated near $p_2=0.1$ and
even decrease for higher $p_{2}$. In contrast, the results obtained with our model agree well with those in \cite{Bertrand2011} for a much larger range. This can be well explained with Eq. \ref{Eq:P3}. In the range of $p_2<0.04$, Eq. \ref{Eq:P3} also gives the scaling $I_{2}^{n_{2}}$ since $p_{1}^{|n_{1}|}p_{2}^{|n_{2}|}$ collapses to $I_{2}^{|n_{2}|}$ when $I_{2}\ll I_{1}$. For $p_2>0.04$, the contribution of the channel $\Omega(n_{1},n_{2})$ reaches its maximum value at $p_{2}=|n_{2}|/(|n_{1}|+|n_{2}|)$ according to Eq. \ref{Eq:P3}. This well explains the saturation effect and the fact that the channel with smaller $n_{2}$ saturates earlier. With the intensity $I_{2}$ further increased ($p_{2}>0.1$), the total intensity increases obviously. As a result, the factor $\sigma(n_{1},n_{2})$
can not be approximated as constant and the harmonic yields obtained from \cite{Bertrand2011} diverge slowly from those obtained from our model. The above results suggest that the HHG in the noncollinear two-color field can be understood essentially from the photon channel perspective and our model provide a more complete and comprehensive insight of the power scaling law than that in \cite{Bertrand2011}. It is also worthy noting that the very recent experiment about HHG with bicicular laser pulses has also been explained based on our model \cite{KM2017}. In addition, our model also has potential to be extended to describe the photon-like features in other processes, such as terahertz generation \cite{vved2016} and HHG in non-pure vortex beam \cite{Laura2016}. For example, the generated terahertz $\Delta\omega=b\omega_2-a\omega_1$ can also be explained in term of photon channel $\Omega(-a,b)$. Our model gives the same scaling laws $P(\Omega(-a,b))\propto (\frac{I_1}{I_1+I_2})^a(\frac{I_2}{I_1+I_2})^b\approx(\frac{I_2}{I_1})^b$ at small $I_{2}$ as in \cite{vved2016}.



In conclusion, a photon channel perspective of HHG is established by quantizing both the driving laser and high harmonics. Our results indicate that the channel competition and power scaling law follows the similar behavior of the nonlinear optical wave mixing. It well explains the HHG in the two-color field and also explains the experimental results in \cite{Bertrand2011} beyond the case of $I_{2}\ll I_{1}$. Our model from the photon channel perspective provides a quantitative approach and useful tool to investigate the quantized photon features in HHG. 


This work was supported by NNSFC (Nos. 11422435, 11234004, 11334009, 11404123, 11425414 and 61475055) and the National Key program for S$\&$T Research and Development (No. 2016YFA0401100).

\end{document}